\documentclass[preprint,12pt]{elsarticle}

\usepackage{amssymb}
\usepackage{float}

\biboptions{numbers,sort&compress}
\begin{document}

\begin{frontmatter}



\title{Emissivity Prediction of Functionalized Surfaces Using Artificial Intelligence}


\author[inst1]{Greg Acosta}
\author[inst2]{Andrew Reicks}
\author[inst1]{Miguel Moreno}

\affiliation[inst1]{organization={Mechanical \& Materials Engineering Department},
            addressline={University of Nebraska-Lincoln}, 
            city={Lincoln},
            postcode={68588}, 
            state={NE},
            country={USA}}
            
\affiliation[inst2]{organization={Electrical \& Computer Engineering Department},
            addressline={University of Nebraska-Lincoln}, 
            city={Lincoln},
            postcode={68588}, 
            state={NE},
            country={USA}}
            
\affiliation[inst3]{organization={Harris Orthopaedics Laboratory, Department of Orthopaedic Surgery},
            addressline={Massachusetts General Hospital}, 
            city={Boston},
            postcode={02114}, 
            state={MA},
            country={USA}}
            
\affiliation[inst4]{organization={Department of Orthopaedic Surgery},
            addressline={Harvard Medical School}, 
            city={Boston},
            postcode={02115}, 
            state={MA},
            country={USA}}

\author[inst3,inst4]{Alireza Borjali}
\author[inst2]{Craig Zuhlke}
\author[inst1]{Mohammad Ghashami}

\begin{abstract}

The radiative response of any object is governed by a surface parameter known as emissivity. Tuning the emissivity of surfaces has been of great interest in many applications involving thermal radiation such as thermophotovoltaics, thermal management systems, and passive radiative cooling. Although several surface engineering techniques (e.g., surface functionalization) have been pursued to alter the emissivity, there exists a knowledge gap in precisely predicting the emissivity of a surface prior to the modification/fabrication process. Predicting emissivity by a physics-based modeling approach is challenging due to surface's contributing factors, complex interactions and interdependence, and measuring the emissivity requires a tedious procedure for every sample. Thus, a new approach is much-needed to systematically predict the emissivity and expand the applications of thermal radiation. In this work, we demonstrate the immense advantage of employing artificial intelligence (AI) techniques to predict the emissivity of complex surfaces. For this aim, we fabricated 116 bulk aluminum 6061 samples with various surface characteristics using femtosecond laser surface processing (FLSP). A comprehensive dataset was established by collecting surface characteristic data, laser operating parameters, and measured emissivities for all samples. We demonstrated the application of AI in two distinct scenarios. First, the range of emissivity of an unknown sample was shown to be estimated correctly solely based on its 3D surface morphology image. Second, the emissivity of a sample was precisely predicted based on its surface characteristics data and fabrication parameters.
The implementation of the AI techniques resulted in the highly accurate prediction of emissivity by showing excellent agreement with the measurements.
\end{abstract}




\end{frontmatter}


\section{Introduction}
\label{sec:Intro}
Emissivity is an important surface radiative property that dictates the output emissive power (i.e., thermal radiation emitted) of any real surface. Emissivity can be defined as the ratio of the radiation emitted by the surface to the radiation emitted by a blackbody at the same temperature \cite{bergman2011fundamentals}. Thus, tuning the emissivity of real surfaces is of great interest in many engineering applications such as thermal protection systems for aerospace vehicles~\cite{zhang2021high,bird2004development,pidan2005recombination}, passive radiative cooling \cite{li2019selective,raman2014passive}, thermophotovoltaics (TPVs) \cite{kim2019nanostructured,burger2020present}, and thermal management systems \cite{Krishna2018,krishna2019ultraviolet}.

\begin{figure}[htp]
\centering
\includegraphics[width=12.5cm]{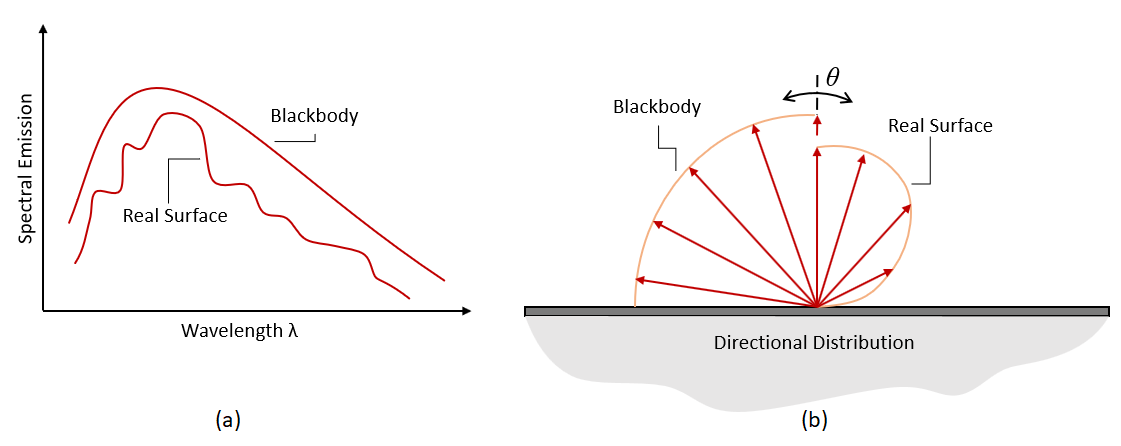}
\caption{Comparison between (a) spectral and (b) directional emission of a blackbody and a real surface.}
\label{fig:DirectionalDistribution}
\end{figure}

Generally, emissivity is a strong function of surface temperature, and depends on the radiation direction and wavelength. Whereas blackbody emission is diffusive and solely depends on a given wavelength and temperature. The dependence of emissivity on wavelength and direction for a real surface can be seen in Fig.~\ref{fig:DirectionalDistribution} as compared to a blackbody.
Now, the spectral hemispherical emissivity can be defined as the net contribution of the spectral, directional emissivity $\varepsilon_{\lambda,\theta}(\lambda,\theta,T)$, across all angular directions of radiation.
\begin{equation}
\label{eq2}
    \varepsilon_\lambda(\lambda,T)=2 \int_{0}^{\pi/2} \varepsilon_{\lambda,\theta}(\lambda,\theta,T) cos \theta sin \theta \,d\theta
\end{equation}
Here, $\lambda$ is the wavelength of radiation, $\theta$ represents the angular direction of the radiation and $T$ the surface temperature. As in most of engineering applications, it is assumed here that the emissivity does not depend on the azimuthal angle, therefore, the total hemispherical emissivity, which represents an average over all possible directions and wavelengths is,
\begin{equation}
    \varepsilon_{}(T)=\frac{\int_{0}^{\infty} \varepsilon_\lambda(\lambda,T)E_{\lambda,b}(\lambda,T)\,d\lambda}{E_b(T)}
\end{equation}
 where $E_{\lambda,b}(\lambda,T)$ is the emissive power of a blackbody at a particular wavelength and ${E_b(T)}$ is the total emissive power of a blackbody over all possible wavelengths



In order to alter the radiative response of a surface, several methods have been suggested in the literature, such as applying coatings and paints on the surface, fabricating metamaterials, or functionalizing surfaces. Coating is a commonly used technique where impurities are introduced as doping agents into an extremely pure surface to modify its electrical and optical properties \cite{he2009high}. For instance, a higher impurity (i.e., doping) concentration can lead to more free electrons within the band, causing more radiation absorption by moving the electrons from a lower-energy level to an empty higher-energy one. Coatings are extensively utilized in TPVs to improve the emissivity of thermal emitters that are otherwise limited \cite{cockeram1999development}. Similar to coatings, paints can also be easily applied to surfaces to enhance the emissivity of different materials. Some applications include radiation pyrometry in turbines \cite{lempereur2008surface, manara2017long}, heating and cooling of buildings \cite{simpson2019thermal, fantucci2019investigating} and thermal imaging of surfaces \cite{manickavasagan2006thermal,brandt2008emissivity}. Even though coatings and paints are eminently suitable for application to curved surfaces or large surface areas, they are limited by their susceptibility to lamination and wear due to variation of environmental conditions. In a different approach, the fabrication of metamaterials has been pursued to obtain surfaces with desired spectral (and directional) emissivity at different bandwidths and temperatures \cite{liu2011taming,liu2016thermochromic}. Metamaterials can exhibit wavelength-selective properties by carefully fabricating sub-wavelength nanostructures on the surface. These nanostructures can enhance the localized electric field in the surface proximity resulting in high absorption or reflection peaks, thus enhancing \cite{baranov2019nanophotonic} or lowering \cite{hu2021thermal} the emissive power of the surface. Although metamaterials offer unique capabilities in precise tuning of the directional and spectral radiative response, they require costly and high-accuracy fabrication techniques that are merely justifiable for small surface areas. Moreover, metamaterials are mostly limited to a narrow spectral range compared to coatings and paints. An alternative way to tune the emissivity of larger surfaces can be achieved by performing surface functionalization. Surface functionalization is a process that involves the combination of physical (e.g., texturing/patterning) and chemical modification of surfaces \cite{florian2020surface}. One way to produce functionalized surfaces is by using femtosecond laser surface processing (FLSP), a unique process to directly modify the surface morphology of almost any material. Studies have demonstrated that laser processing can be used to modify how surfaces interact with electromagnetic waves to improve their broadband absorption or emission \cite{fan2013rapid, chen2020multi,tang2012nanosecond}. In FLSP, the material's surface properties are modified by the formation of unique micro-/nano-scale surface features, and microstructure and surface chemistry changes that alter the optical properties of the surface \cite{singh2006femtosecond}. The characteristics of these micro-/nano-scale structures can be directly correlated to the FLSP operating parameters such as laser fluence (i.e., energy per surface area) and laser pulse count \cite{ou2016fluence}.
FLSP surfaces have greater permanency and durability when compared to coatings and paints, two key properties that are crucially important for applications in extreme environments. Compared to metamaterials, FLSP surfaces demonstrate remarkably wider bandwidth and lower fabrication complexity \cite{Reicks}.

Even though the above methods show some of the great technological advancements in engineering surfaces to alter the emissivity, there still remains a significant challenge in predicting the outcome of these processes. In other words, it is difficult to exactly know what emissivity to expect from the modified surfaces before performing the fabrication/modification process. For instance, to fabricate a sample with a desired emissivity using FLSP, one needs to precisely know what surface features (or alterations) are needed to provide the right result. In principle, this is feasible by having a fundamental understanding of how various surface patterns and microstructures affect the interaction of electromagnetic waves with the surface. Therefore, it is imperative to develop a deeper understanding of this interaction, in order to systematically engineer surfaces with optimal emissive power, and expand the capabilities of thermal radiation applications. However, trying to correlate the geometrical attributes of the surface structures to the measured emissivity via physics-based or model-driven approaches can be a cumbersome task due to the multifaceted nature of the problem. For example, the laser parameters in FLSP control the characteristics similar to the parameters used to apply the coatings. Coatings are generally flat films where the thickness and doping concentration define the resultant emissivity. In contrast, the laser processed surfaces are textured, where the morphology of the micro-/nano-scale features, surface chemistry and subsurface microstructure (porosity, nanoparticle layering, oxidation thickness, etc.) induce the radiative response of the surface. Due to this high dependence of interconnected variables in emissivity, a new approach is needed to facilitate obtaining a predictive model. Artificial intelligence (AI) techniques can be employed to act as facilitators to capture the unique features of the surface and correlate the operating parameters to the obtained emissivity. While classical physical modeling requires an in-depth knowledge of the phenomena and often involves some simplifying assumptions to develop the closed-form mathematical models, AI-based data-driven models of physical processes in various fields have demonstrated great potential to accurately predict physical properties, especially when the physics-based analytical or traditional statistical models are not easily available \cite{sizemore2020application,wuestmachine,kim2018smart,jurkovic2018comparison}. By exploiting the unique capabilities of AI, recent studies have developed predictive data-driven models of physical phenomena that otherwise might be challenging to obtain via analytical methods or will be computationally expensive. For example, Borjali $\textit{et al.}$ \cite{Borjali2019}, developed a data-driven model for predicting wear rate of orthopaedic polyethylene as a function of the wear experiment's parameters such as velocity and contact area. In a different application, Xiong $\textit{et al.}$ \cite{xiong2020machine} developed a predictive model using AI to understand how the shear and bulk modulus of new bulk metallic glasses (BMGs) are affected by alloy composition.
In an effort to predict the surface roughness of additive manufactured Ti-6Al-4V, Akhil $\textit{et al.}$ \cite{akhil2020image} used AI to extract texture parameters from scanning electron microscopy (SEM) images and together with the measured sample surface roughness, developed a predictive model.

Beside the massive advantage in developing predictive data-driven models, AI techniques after successful training can be implemented for inverse design and optimization problems in different engineering disciplines. For instance, in nanophotonics, AI is used to optimize the subwavelength geometrical features of photonic structures and hence the optical response of the material \cite{so2020deep}. Peurifoy $\textit{et al.}$ \cite{peurifoy2018nanophotonic} used deep neural networks (DNNs) to first predict the light scattering of a multilayered core-shell nanoparticle. Once the DNN was trained to successfully predict the phenomenon, it was used to optimize the total number of layers and their thicknesses required to achieve a desired optical response. Similarly, So $\textit{et al.}$ \cite{so2019simultaneous} used a special DNN architecture to simultaneously design and output the optimal material and layer thickness of spherical three-layered nanoparticles based on a set of desired electric and magnetic dipole resonances as the input. In another application, Liu $\textit{et al.}$ \cite{liu2018generative} implemented an ensemble of convolutional neural networks (CNNs) to generate optimal surface patterns for structured metasurfaces, where the input to the network was a desired spectral transmittance distribution. Lastly, Garcia $\textit{et al.}$ \cite{garcia2021deep} demonstrated how deep learning techniques can be implemented for the modeling and inverse design of radiative heat transfer phenomena in various systems including hyperbolic metamaterials, passive radiative cooling in photonic-crystals, and emissive power of subwavelength objects. 

The aforementioned studies are just a handful of many diverse research projects where the applications of AI have proven to be of significant importance in the analysis of data, prediction of physical properties and inverse design of physical phenomena. In this study, we demonstrate how AI techniques can be employed to successfully predict the total hemispherical emissivity of functionalized aluminum 6061 alloy samples. The FLSP technique was chosen to functionalize the samples' surfaces because of its full control over the fabrication parameters, high repeatability, scalability, and ease of fabrication.
The surface characteristics and morphology of all samples are extracted using laser scanning confocal microscopy (LSCM) and the total hemispherical emissivity is measured by infrared imaging, as will be discussed in detail later. By collecting all the surface parameters and emissivity of these samples, a comprehensive dataset is built that serves to train, validate and test the predictive data-driven models. The predictive data-driven model correlates the surface features (extracted from LSCM images) and the laser processing parameters such as peak fluence and pulse count, to the total hemispherical emissivity. In the following sections, the sample fabrication, preparation and characterization processes will be discussed in detail, followed by the AI implementation process and discussion of results. This data-driven modeling approach opens new paradigms for predicting physical phenomena that are highly dependent on complex interactions.

\section{Sample Preparation and Fabrication via FLSP}
To directly modify the surface properties of the bulk aluminum alloy 6061 samples in a well-controlled manner, the FLSP technique has been employed in this work.
FLSP has many advantages over other surface functionalization methods: it results in a fully functionalized surface in a single processing step; it is a scalable process; it involves the creation of hierarchical micro-/nano-scale surface features composed of the original material, making the surface highly permanent; it leads to modification of the original surface without the net addition of mass; and, it results in minimized heat affected zone, so the surface can be modified without altering the bulk properties of the materials \cite{le2002comparison}. More importantly, FLSP surfaces can produce omnidirectional emissivity due to high absorption at large incident angles , which is very difficult to achieve via coatings, paints, or metamaterials \cite{huang2015blackening,Reicks}. FLSP can form quasi-periodic patterns of self-organized microstructures. The geometrical structure of these permanent surface features mostly resembles micro-/nano-scale mounds or pyramids, coated by a thin layer of redeposited nanoparticles \cite{Reicks}. In addition to the formation of the multiscale patterns on the surface, the surface chemistry and subsurface microstructure will be altered by FLSP, leading to very unique surface properties for each sample. By adjusting the processing parameters of FLSP, such as fluence, pulse count, and the atmospheric environment, one can directly control the resultant surface morphology and chemistry. 

\begin{figure}[h!]
\centering
\includegraphics[width=11.5cm]{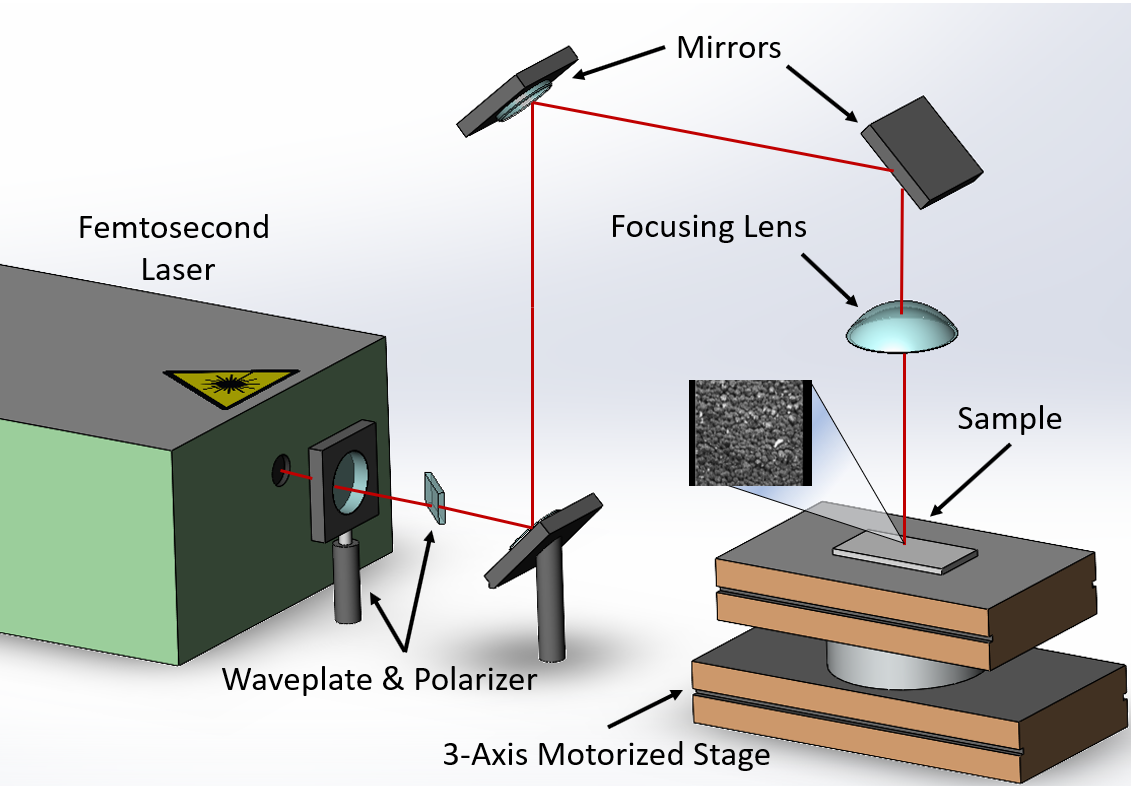}

\caption{Schematic illustration of the femtosecond laser surface processing setup. The samples are mounted on 3-axis motorized stages that control the processing pattern of the laser. Since the beam diameter is small compared to the size of the processed area, a rastering pattern is used to effectively cover the surface.}
\label{fig:FLSP_setup.png}
\end{figure}

Prior to applying FLSP, the samples were cleaned to remove any contamination by wetting with ethanol and allowed to dry. Afterwards, the samples were placed on a motorized stage within an open air environment, where the surface processing occurs. A typical setup to apply FLSP is shown in Fig.~\ref{fig:FLSP_setup.png}, consisting of a femtosecond laser system, beam delivery and focusing optics, and motorized 3D stages exposed to an open air  environment, where samples are placed. The laser used was a Coherent Inc. Astrella Ti:sapphire laser system that produces 6 mJ, 35 fs pulses at a 1 kHz repetition rate, with a central wavelength of 800 nm. The pulses were focused onto the sample surface using a 150 nm focal length plano-convex lens.  
The laser output is quantified via two laser processing parameters known as the fluence and pulse count. The fluence values given are the peak fluence, which is the fluence at the peak of the Gaussian distribution of the focused pulses. Peak fluence $(F_p)$ is the energy per unit area $(J/cm^2)$, and is defined as,
\begin{equation}
\label{fluence}
    F_p = \frac{8P}{2 \pi \omega^2 R}
\end{equation}
where, $P$ is the average power, $\omega$ is the $1/e^2$ beam radius, and $R$ is the repetition rate of the laser. In order to process an area larger than that of the beam, a raster scanning pattern is utilized.  The laser is used to scan a line in the x direction and then the motorized stage is stepped over in the y direction. The step distance between line scans is referred to as the pitch, $p$. The pulse count $(PC)$ considers the overlap in the pitch and scan directions to calculate the number of pulses incident at each point on the surface and is defined as, 
\begin{equation}
\label{pulsecount}
    PC =  \frac{R}{2 \omega v}(\frac{\omega}{p})
\end{equation}
where $v$ is the stage velocity, and $R$ is the repetition rate of the laser. Optimizing the fluence and pulse count is a crucial step as they dictate the shape and periodicity of the microstructures, and also the thickness of the oxide layer forming on the surface.

\begin{figure}[h!]
\centering
\includegraphics[width=13.5cm]{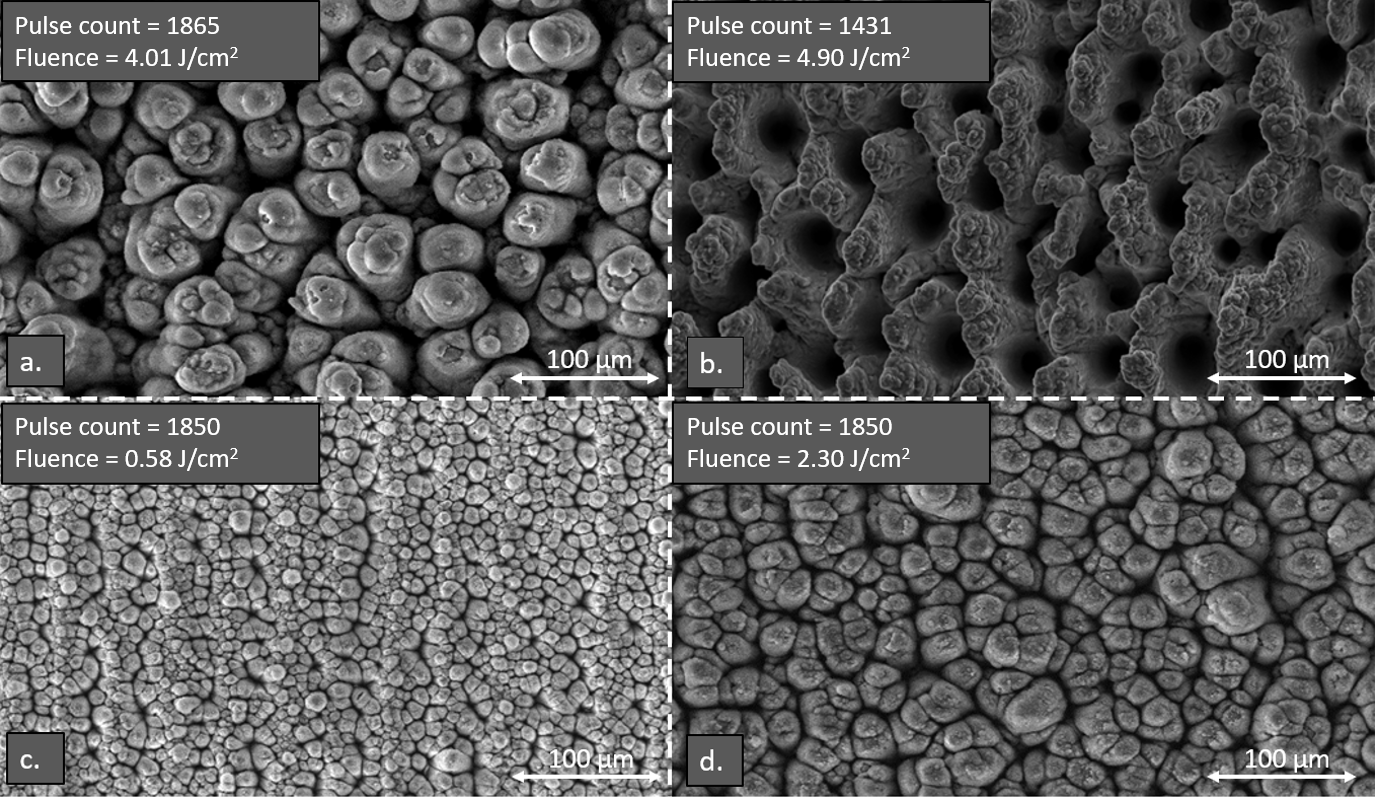}
\caption{Surface SEM images of the aluminium samples with quasi-periodic microstructures produced by FLSP. The laser parameters of pulse count and fluence are shown on each image. The resultant measured hemispherical emissivities are (a) $\varepsilon_h$ = 0.926, (b) $\varepsilon_h=0.865$, (c)  $\varepsilon_h=0.781$, and (d) $\varepsilon_h=0.926$.}
\label{fig:SEM_Images_final.png}
\end{figure}

In order to produce the 116 samples used  for this study, the fluence was varied between 0.06 and 5.5 $(J/cm^2)$ and the pulse counts between 270 and 14000. Fig.~\ref{fig:SEM_Images_final.png} shows surface SEM images of four of the aluminum alloy 6061 samples after the FLSP is performed. From Fig.~\ref{fig:SEM_Images_final.png}, the formation of quasi-periodic self-organized microstructures can be observed. Note the different morphology for the micro-/nano-scale FLSP surface features that develop for the different fluence and pulse count values.

\section{Surface Characterization and Emissivity Measurement}
\begin{figure}[htp]
\centering
\includegraphics[width=13.5cm]{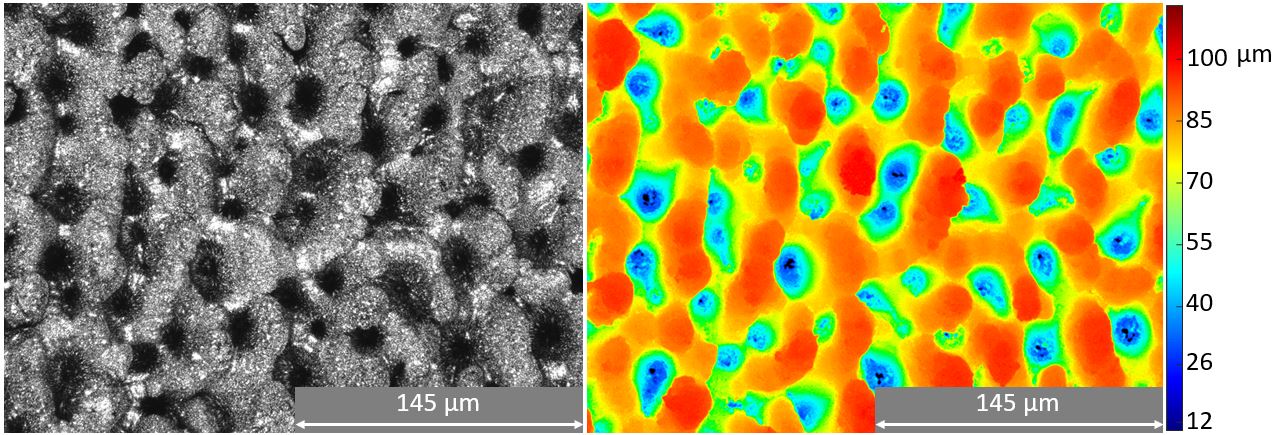}
\caption{(Left) 3D LSCM image of the FLSP aluminum sample captured at 50$\times$ magnification, (Right) 3D topographic map and height measurement of the same sample. This sample was produced with a pulse count of 490 and peak fluence of 2.01 $J/cm^2$.}
\label{fig:SAMPLE.png}
\end{figure}

In order to fully characterize the surface features of the FLSP samples, a LSCM (Keyence VK-X200K) with 500 nanometer Z-axis resolution was used with a 50$\times$ objective to capture the 3D topography at three different areas of each sample. Three scanning areas were chosen to account for the potential variations of the patterns along the surface. The average height $R_z$ and the average roughness $R_a$ for each surface were extracted from these images using VK-Analyzer software. Based on these measured properties, the average skewness and the average kurtosis were calculated as well.
Fig.~\ref{fig:SAMPLE.png} shows the comparison between the optical image and laser scanning 3D height map of a functionalized aluminum sample captured with the LSCM. The red shaded regions in the height-filtered image represent higher elevation or peaks, whereas the blue shaded regions represent lower elevation or valleys. 
Another important piece of data that needs to be extracted from the height-filtered images is the density distribution (i.e., number of mounds per unit surface area) of the microstructures on each surface. To accomplish this, the Fast Fourier Transform (FFT) was implemented using MATLAB to find the strongest or the most frequent pattern in the LSCM images. Using this image processing technique, the dominant frequency in a 2D plane was found. In some images, however, a strong periodicity may be observed only in one direction. For such cases, the periodicity (or spacing) is usually assumed to be equal in both directions. The density distribution can be then determined by knowing the periodicity of the patterns and the scale of each LSCM image.

\begin{figure}[H]
\centering
\includegraphics[width=11.5cm]{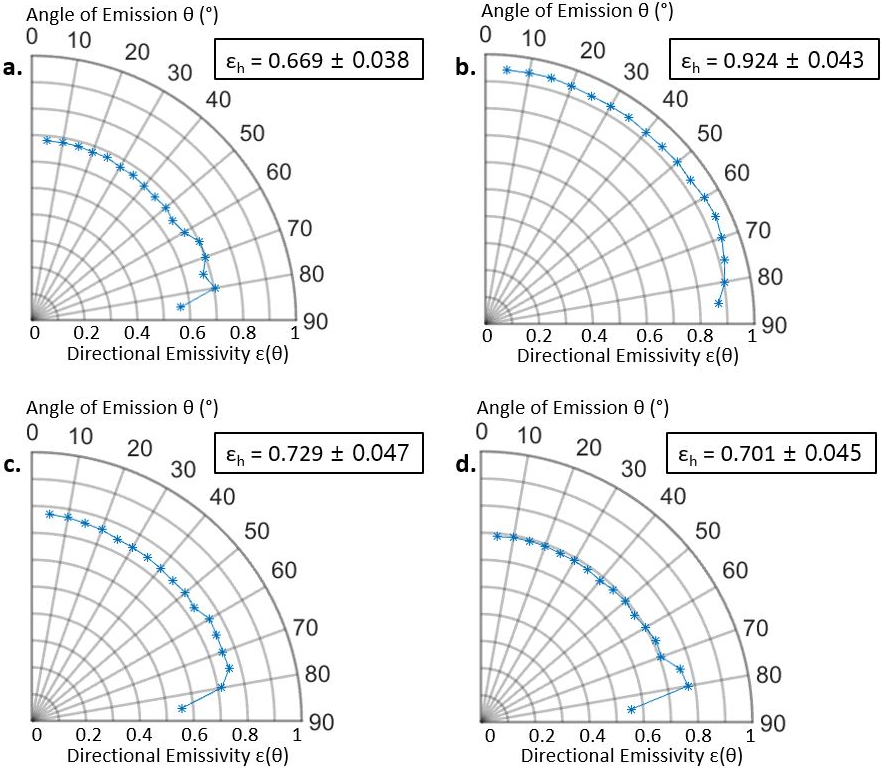}
\caption{Measured directional distribution of emissivity and hemispherical emissivity for four different FLSP samples: (a) Pulse count of 490 and fluence of 2.01 $J/cm^2$. (b) Pulse count of 6875 and fluence of 5.5 $J/cm^2$. (c) Pulse count of 762 and fluence of 2.10 $J/cm^2$. (d) Pulse count of 490 and fluence of 2.62 $J/cm^2$.}
\label{fig:Em_results.png}
\end{figure}

Now, to calculate the hemispherical emissivity of a surface, the directional emissivity is necessary. To this aim, we utilized a FLIR A655sc thermal imaging camera with a spectral range of 7.5 to 14 $\mu$m to measure the surface temperature of both a calibrated source and the sample uniformly heated to 60 $^\circ C$ at different angles from 0° to 85° with increments of 5°. By establishing an energy balance over the detector of the thermal camera we can correlate the energy of the detector, $E_d$, to the directional emissivity of a sample, $\varepsilon$, 
\begin{equation}
\label{energybalance}
    E_d = \varepsilon(\theta) E_s + [1 - \varepsilon(\theta)] E_b
\end{equation}
where $E_s$ is the energy emitted from the sample, and $E_b$ is the background energy from the measurement. The energies of the detector, sample, and the background are related to their temperatures via the Stefan-Boltzmann law, $E = \sigma T^4$. By utilizing the measured temperature of a calibrated source and its known emissivity, the temperature and thus the energy of the detector can be found. In this work, black polyvinyl chloride electrical tape was used as the calibrated source with a known hemispherical, and directional emissivity measured via Surface Optics SOC-100 reflection based instrument. Once the energy of the detector is known, the directional emissivity of the unknown sample can be obtained from Eq.~\ref{energybalance2}, as the ratio of the energies of the detector to that of the unknown sample minus some small background contribution,  \cite{Reicks}, 

\begin{equation}
\label{energybalance2}
   \varepsilon(\theta) = \frac{E_d - \sigma T_b^4}{\sigma T_s^4 - \sigma T_b^4} 
\end{equation}
It is noteworthy to express that the spectral dependence in Eq.~\ref{eq2} is not considered when approximating the hemispherical emissivity, since the thermal imaging camera operates in the wavelength range of 7.5 to 14 $\mu$m. This, in effect, averages the measured emissivity with respect to the wavelength. Since the measured directional emissivities are for discrete angles, a numerical integration must be employed to calculate the total hemispherical emissivity based on Eq.~\ref{eq2}. For the approximation, the average between the rectangular and trapezoidal numerical integration was used. The difference between the approximate error of both methods yields the overall numerical uncertainty in calculating the total hemispherical emissivity \cite{Reicks}. Fig.~\ref{fig:Em_results.png} shows the measured directional and hemispherical emissivity of four different samples. The results depict that the magnitudes of the peak fluence and pulse count play decisive roles in the outcome of the overall emissivity. This is expected since the oxide layer thickness, structure height, and the periodicity of the microstructures depend on the magnitudes of the peak fluence and pulse count.
A beam with lower fluence will result in finer structures, whereas a beam with higher fluence will result in coarser structures. In addition, at higher fluences the oxide layer is not uniform, resulting in a lower emissivity. Hence, there exists an optimal peak fluence and pulse count that will yield an optimal emissivity.

\section{Data-Driven Prediction of Total Hemispherical Emissivity}
In this study, 116 different samples were fabricated, characterized, and tested to build the study's dataset. The dataset consists of ($i$) laser operating parameters: pulse count $P_c$, fluence $F_p$, and total fluence $F_t$; ($ii$)  surface characteristics data: average height $R_z$, average roughness $R_a$, average skewness $S_k$, average kurtosis $K_u$, mound surface area to planar area ratio $S_A$, and mound concentration (or density) $D$; and ($iii$) the measured total hemispherical emissivity, $\varepsilon_h$.
In addition, we collected 250 LSCM images from these samples. Multiple images were taken from different areas of each sample to create the image dataset. 

The capabilities of AI in emissivity prediction of a surface were studied for two different scenarios. First, we wanted to know if AI can be employed to predict the (range of) emissivity of a new sample just based on its 3D LSCM image without providing any other information about its surface characteristics or fabrication parameters. In other words, if we have a 3D surface morphology image of a sample without knowing anything else about it, can we estimate its expected range of emissivity? The successful accomplishment of this step is particularly advantageous for cases where an approximate surface radiative property is needed, and the only available data is a 3D morphology image of the sample and prior knowledge of the processing conditions.
In the second scenario, we wanted to go one step further to know if we can  precisely predict the actual emissivity of a new FLSP sample based on its surface characteristics data and laser operating parameters. Attaining such a model will provide us with a powerful tool that obviates the need for the costly procedures to measuring the surface radiative properties of a sample and by reducing the parameter studies needed to produce a certain emissivity.  
\begin{figure}[b!]
\centering
\includegraphics[width=13.5cm]{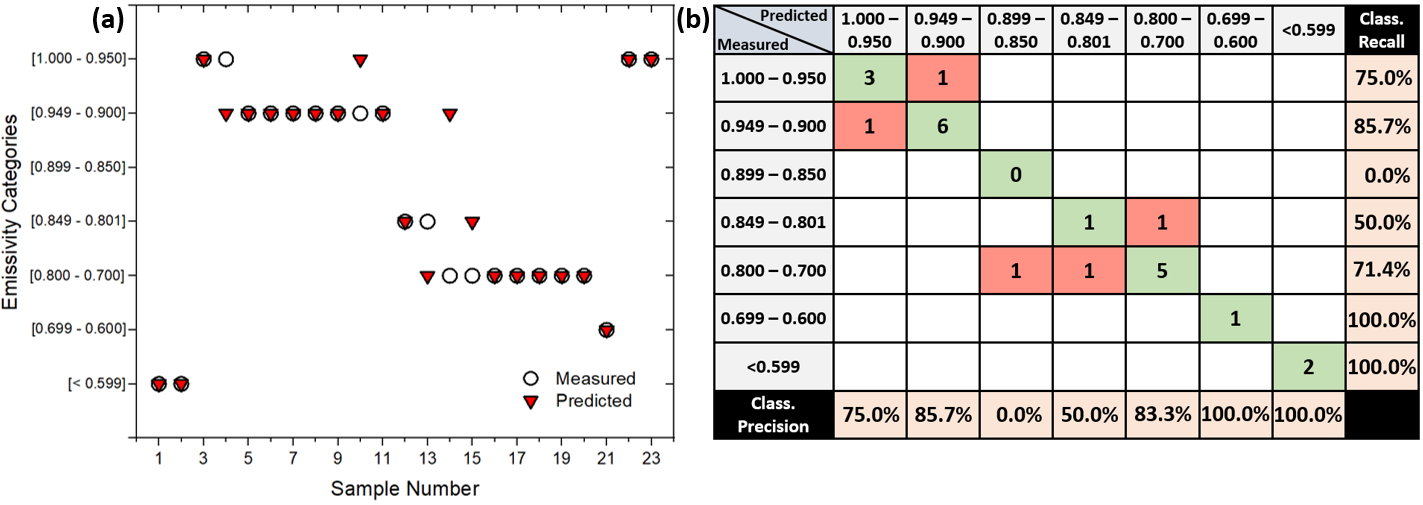}
\caption{(a) Classification results for $\textit{Model 1}$, where the CNN's extracted features where used as inputs. Here, an error is predominantly present between categories closed to each other, suggesting that some samples have similar surface features, thus confusing the CNN feature extractor. (b) Confusion matrix depicting the classification error in our $\textit{Model 1}$. A green shaded area represents a positive (correct) response and a red shaded area represents a negative (incorrect) response. From the positive and negative responses, the classification precision (i.e., the ratio of correct predictions to the total relevant samples) and recall (i.e., the ratio of correct predictions to the total predicted cases)  were calculated for each emissivity range. $\textit{Model 1}$ was tested on a total of 23 images.}
\label{fig:DL_results.png}
\end{figure}
To test the feasibility of the first scenario, we used the measured hemispherical emissivity as the ground truth and divided the entire image dataset into seven categories based on the measured emissivities. These emissivity categories ranged between 0.599 and 1, as illustrated in Fig.~\ref{fig:DL_results.png}. We used a split validation method throughout this study by randomly dividing the dataset into training, validation, and final test subsets with an 80:10:10 split ratio. We developed an AI model, $\textit{Model 1}$, that solely used the LSCM images. As a precursor to this model, we implemented transfer learning by adopting VGG16 CNN architecture that was pre-trained on the ImageNet dataset. We used this CNN as a feature extractor for analyzing the LSCM images and developed a neural network (NN) to categorize the  features extracted from the images into seven emissivity categories.
Fig.~\ref{fig:DL_results.png} shows the results for $\textit{Model 1}$. It can be seen from Fig.~\ref{fig:DL_results.png}(a) that the model performed well and there is a great match for the majority of our test samples. It can be noted that for samples 4, 10, 13, 14, and 15, there is a mismatch in the classification process. From the confusion matrix in Fig.~\ref{fig:DL_results.png}(b), it can be observed that the error is predominantly present between categories close to each other, except for sample number 14. This mismatch can be attributed to the samples having similar surface features between the adjacent categories, thus confusing the CNN. It should be noted that by increasing the size and diversity of the image dataset, the model could get trained much better which would significantly improve its classification ability and mitigate such confusion between neighboring categories. With a larger dataset, the width of the selected categories could be narrower to improve the accuracy of the estimations.

\begin{figure}[H]
\centering
\includegraphics[width=13.5cm]{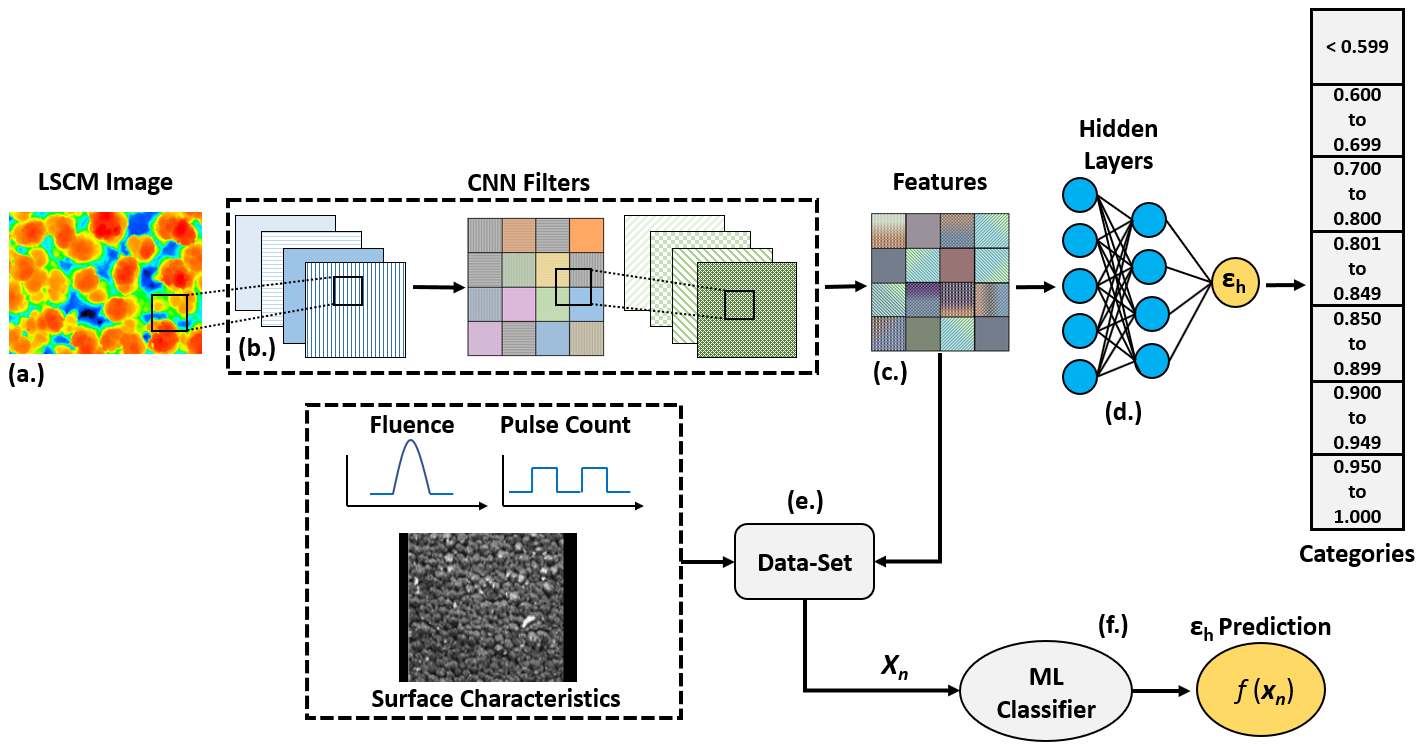}
\caption{General description of the AI architecture implemented to predict the emissivity of the FLSP samples. (a) Here, each captured LSCM image serves as an input to our CNN architecture. (b) The initial LSCM images are processed one-by-one, through a series of convolutional filters that produce a feature vector. (c) Final feature vector then serves as an input into (d) $\textit{Model 1}$,  a general neural network that classifies each image into seven different emissivity categories. (e) A comprehensive dataset is built consisting of the laser operating parameters, measured surface characteristics and the detected features of the LSCM images. Then this data is fed into (f), our $\textit{Model 2}$ that predicts the emissivity as a function of the input parameters.}
\label{fig:Esemble_AI.png}
\end{figure}

Now, to demonstrate the application of AI in the second scenario, we developed $\textit{Model 2}$. For this model, we combined features extracted from the LSCM images with laser operating parameters and surface characteristics to build a comprehensive dataset. Fig.~\ref{fig:Esemble_AI.png} illustrates the AI architecture that was implemented by combining the image processing through deep-learning with several machine learning classifiers to predict the emissivity. These classifiers included k-nearest neighbor (kNN), artificial neural network (ANN), generalized linear model (GLM), W-M5P, and decision tree (DT). Detailed explanation of these classifiers can be found elsewhere and will not be repeated here \cite{hastie1996discriminant, dreiseitl2002logistic, kilicc2013linear, rokach2007data}. We used the training subset (80\% of our dataset) to develop this model. Hyper-parameters were optimized iteratively on the validation subset using a grid search strategy. The final models were tested on the test subset that was isolated from the training and validation process. The performance of each model on the test subset is summarized and reported in Table~\ref{table:2}, where root mean square error ($RMSE$), absolute error ($ABS$), and the coefficient of determination ($R^2$) are tabulated for comparison.

\begin{figure}[H]
\centering
\includegraphics[width=13.7cm]{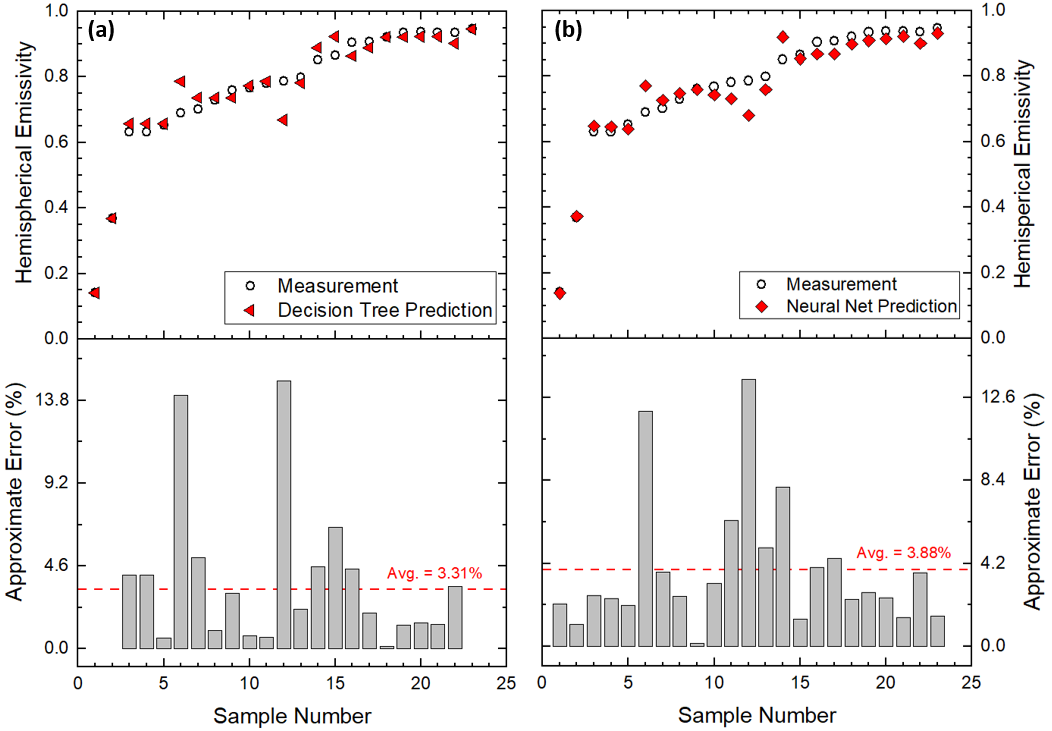}
\caption{Prediction results for our $\textit{Model 2}$ using: (a) a decision tree (DT) classifier with an approximate error of 3.31\%, and (b) an artificial neural network (ANN) classifier with an approximate error of 3.88\%.}
\label{fig:DT_Results.png}
\end{figure}

\begin{table}[htp]
\centering
\caption{Prediction performance of different machine-learning models}
\label{table:2}
\begin{tabular}{||c|c| c| c||} 
 \hline
 Model & $RMSE$ & $ABS$ & $R^2$ \\ [0.5ex] 
 \hline\hline
 W-M5P & 0.049  & 0.035 $\pm$ 0.034 & 0.966 \\ 
 GLM & 0.066 & 0.055 $\pm$ 0.038 & 0.952 \\
 kNN & 0.047  & 0.036 $\pm$ 0.030 & 0.978 \\
 DT & 0.039  & 0.026 $\pm$ 0.029 & 0.979 \\
 ANN & 0.039 & 0.030 $\pm$ 0.025 & 0.980 \\ [1ex] 
 \hline
\end{tabular}
\end{table}
From table~\ref{table:2}, it can be observed that DT and ANN outperformed the rest of the classifiers. The obtained predictions using DT and ANN classifiers are illustrated in Fig.~\ref{fig:DT_Results.png}. From the results, it can be observed that both classifiers performed well with great match between the predictions and the measurements, where DT and ANN yielded an average approximate error of 3.31\% and 3.88\%, respectively.     
\noindent It can also be observed that for the samples number 6 and 12, there is a large deviation from the true value of measured emissivity in both DT and ANN methods. This can potentially be attributed to the error originated in feature extraction from the CNN in $\textit{Model 1}$ that has propagated into $\textit{Model 2}$. This type of error can be addressed by having a larger LSCM image-set and hence a better performance of the CNN.

\section{Conclusion}
In this study, we have demonstrated the immense advantage of applying AI techniques to predict the emissivity of complex surfaces. For this purpose, we fabricated 116 bulk aluminum alloy 6061 samples using FLSP. The distinct microstructures formed on these surfaces altered the emissivity of the samples. We extracted the surface morphology for all samples and collected 250 3D LSCM images by performing rigorous surface characterization. The directional emissivity of the samples was measured using a thermal imaging camera with a spectral range of 7.5 to 14 $\mu$m. The directional emissivity was then numerically integrated to calculate the total hemispherical emissivity of samples.

We demonstrated the application of AI for emissivity prediction in two different cases. In case 1, we showed that the emissivity range for a given surface could be approximated merely based on its 3D morphology image. We pre-trained a CNN that served as a feature extractor on our training image dataset and developed an ANN to classify the test samples into seven emissivity categories. The obtained results revealed the great advantage of AI-based methods in estimating the emissivity by image processing.
For case 2, we demonstrated that the combination of deep learning and machine learning techniques could be implemented to precisely predict the emissivity of a sample by knowing its surface characteristics and fabrication parameters. To accomplish this, several machine-learning classifiers were applied to the dataset where DT and ANN classifiers showed the best performance with approximate errors of 3.31\% and 3.88\%, respectively. Despite a noticeable error for two samples, we were able to accurately predict the total hemispherical emissivity of functionalized aluminum surfaces produced by FLSP. This alternative data-driven approach opens new paradigms for predicting physical phenomena that might otherwise be difficult to predict by classical physics-based modeling.

\section*{Declaration of Competing Interest}
The authors declare that they have no known competing financial interests or personal relationships that could have appeared to
influence the work reported in this paper

\section*{Acknowledgements}
This work was supported in part by the National Aeronautics and Space Administration (NASA) Nebraska Space Grant NNX15AI09H. Manufacturing and characterization analysis were performed at the Nano-Engineering Research Core Facility, University of Nebraska-Lincoln, which is partially funded from the Nebraska Research Initiative funds. We would also like to acknowledge the NASA Nebraska Space Grant Fellowship 4403071026358 and the NASA-EPSCoR Mini-Grant 4403071025314 for their financial support. 

 \bibliographystyle{elsarticle-num} 
 \bibliography{refs_test}





\end{document}